%% file: main.tex
\documentclass[preprint,journal]{vgtc}            

\usepackage{xurl}
\usepackage{soul}
\usepackage[dvipsnames, svgnames]{xcolor}

\newcommand{\osf}{\url{https://osf.io/tn6m2/}}
\newcommand{\liveLink}{\url{https://chicago-pb-probe.netlify.app/}}

\newcommand{\figref}[1]{\hyperref[#1]{Fig.~\ref*{#1}}}
\newcommand{\secref}[1]{\hyperref[#1]{Sec.~\ref*{#1}}}

\newcommand{\etal}{et al.}

\newcommand{\ie}{{i.e.,}}
\newcommand{\eg}{{e.g.,}}

\newcommand{\hlc}[2][yellow]{{%
                  \colorlet{foo}{#1}%
                  \sethlcolor{foo}\hl{#2}}%
}
\definecolor{quoteColor}{HTML}{B3DDF2}

\newcommand\qt[1]{\hlc[quoteColor]{``#1''}}
\newcommand{\pxx}[1]{\textbf{P$_{#1}$}}

\usepackage[normalem]{ulem}
\definecolor{linkColor}{HTML}{257E98}
\setuldepth{Berlin}

\usepackage{tcolorbox}

\definecolor{synthesisColor}{HTML}{f0f0f0}

\tcbuselibrary{breakable}
\tcbset{
  width=0.98\columnwidth,
  halign=justify,
  center,
  breakable,
  colback=synthesisColor    
}

\newcommand{\synthbox}[1]{
\begin{tcolorbox}
  #1
\end{tcolorbox}
}

\definecolor{revisionColor}{HTML}{FDFF32}

\newcommand\revision[1]{#1}

\usepackage{graphicx}
\makeatletter 
\define@key{Gin}{alt}{} 
\makeatother 

\onlineid{1281}

\vgtccategory{Research}

\vgtcpapertype{evaluation}

\title{What Can Interactive Visualization do for \\ Participatory Budgeting in Chicago?}

\author{Alex Kale, Danni Liu, Maria Gabriela Ayala, Harper Schwab, Andrew McNutt}

\authorfooter{
    \item
    Alex Kale is with the University of Chicago. E-mail: kalea@uchicago.edu.
    \item
    Danni Liu is with the University of Chicago. E-mail: danni6@uchicago.edu.
    \item
    Maria Gabriela Ayala is with the University of Chicago. E-mail: mariagabrielaa@uchicago.edu.
    \item
    Harper Schwab
     is with the University of Chicago. E-mail: hwschwab@uchicago.edu.
    \item
    Andrew McNutt is with the University of Washington. E-mail: amcnutt@uw.edu.
}

\abstract{%
    Participatory budgeting (PB) is a democratic approach to allocating municipal spending that has been adopted in many places in recent years, including in Chicago. Current PB voting resembles a ballot where residents are asked which municipal projects, such as school improvements and road repairs, to fund with a limited budget. In this work, we ask how interactive visualization can benefit PB by conducting a design probe-based interview study (N=13) with policy workers and academics with expertise in PB, urban planning, and civic HCI. Our probe explores how graphical elicitation of voter preferences and a dashboard of voting statistics can be incorporated into a realistic PB tool. Through qualitative analysis, we find that visualization creates opportunities for city government to set expectations about budget constraints while also granting their constituents greater freedom to articulate a wider range of preferences. However, using visualization to provide transparency about PB requires efforts to mitigate potential access barriers and mistrust. We call for more visualization professionals to help build civic capacity by working in and studying political systems.
}

\keywords{Visualization, Preference elicitation, Digital democracy}

\CCScatlist{ 
 \CCScat{H.m}{Information Systems}%
{Miscellaneous}
}

\teaser{
    \centering
    \includegraphics[alt={A comic-style teaser figure depicting our design scenario in four panels. Panel 1: Residents face issues that they would like city goverment to address; pictured is a person complaining about traffic. Panel 2: Participatory budgeting (PB) collects these issues; pictured is a person with a clipboard going door-to-door chatting with neighbors. Panel 3: PB gives residents a say in which issues should be funded; pictured is a snapshot of the interactive visualization used to elicit PB voting in our design probe. In this work, we explore the design of how preferences are elicited after topics of interest have been gathered via a computer based design probe. Panel 4: PB potentially increses engagement with government and satisfaction with spending; pictured is the person from panel 1 relieved to see s street filled with bikes rather than cars.}, width=\linewidth]{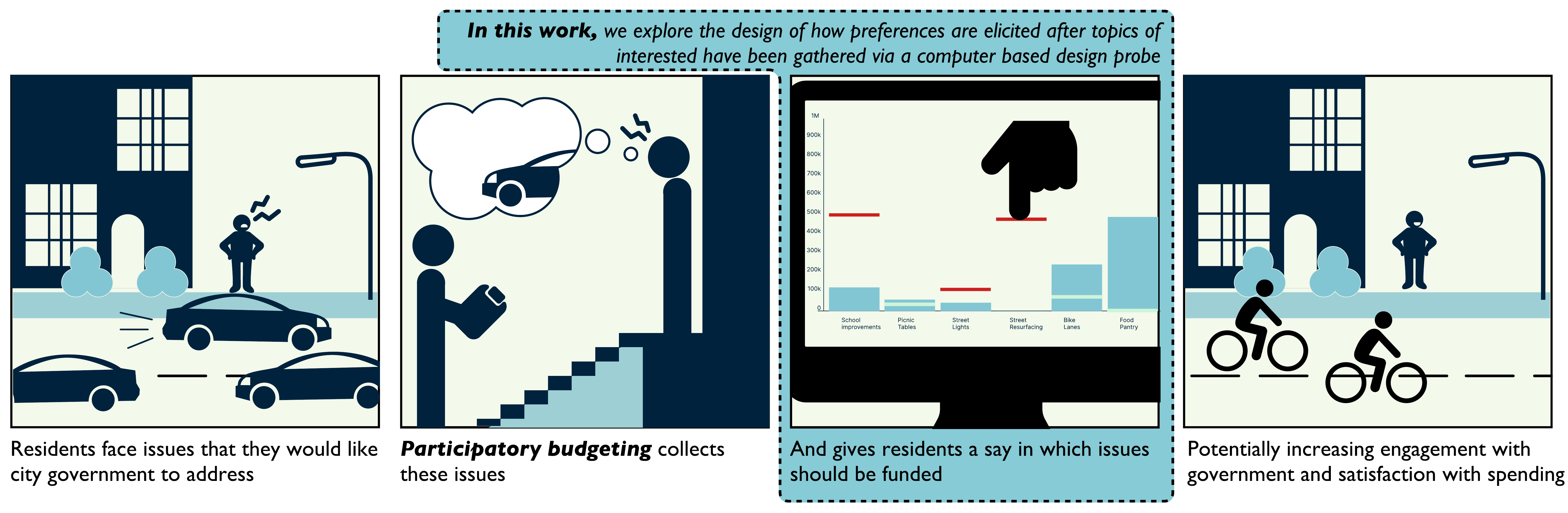}
    \vspace{-2em}
    \caption{
        An illustration of the application scenario for this work, participatory budgeting.
    } 
    \label{fig:teaser}
}

\graphicspath{{figs/}{figures/}{pictures/}{images/}{./}} 

\usepackage{tabu}                      
\usepackage{booktabs}                  
\usepackage{lipsum}                    
\usepackage{mwe}                       

\usepackage{mathptmx}                  

\begin{document}


\firstsection{Introduction}

\maketitle

\input{1_intro.tex}
\input{2_related}
\input{3_probe}
\input{4_method}
\input{5_results}
\input{6_discussion}

\acknowledgments{
We thank the University of Chicago Data Science Institute for funding this work. We also thank our participants for sharing their time and perspective, without which this work would not have been possible.
}

\bibliographystyle{abbrv-doi-hyperref}

\bibliography{0_pb-util}

\appendix 

\end{document}

%% file: 1_intro.tex

Can visualization facilitate social good?
Recent efforts from the visualization community have explored how the presentation and interaction tools that define our field can benefit society.
For instance, visualization of opinions can support deliberation in community meetings~\cite{Jasim2021-CommunityClick,Jasim2021-CommunityPulse,Baumer2022-political}. 
More broadly, interactivity in online articles makes explanations of data and complex processes more engaging and memorable~\cite{Conlen2019-capture,hohman2020communicating,Nguyen2018-BeliefDrivenDJ}, supporting a more informed populace. 
Compared with national issues like presidential elections, local government tends to receive substantially less attention and resources, suggesting opportunities for social impact. 
We ask how the affordances ubiquitous to visualization---such as their propensity to make potentially prosaic topics pertinent---might create avenues for ``empowered participation''~\cite{Fung2004-empowered} in local government.

To answer this question, we consider the case of Chicago's use of participatory budgeting (PB) (\figref{fig:teaser}).
PB is a procedure for allocating municipal funds by direct democracy that has gained traction in many cities in recent years~\cite{Palacin2023-PB-configurations,StanfordPB} (see \secref{sec:pb-chi}) and enjoys support from open-source software platforms~\cite{Consul,Decidim,Holston2016-AppCivist-PB,Menendez-Blanco-2022-PB-Madrid-biking,StanfordPB} (see \secref{sec:pb-tools}).
We take a particular interest in PB because, in addition to being an area of social import that operates on local scale, it represents a real-world collective decision problem with at least two major challenges that visualization might help to address: (1) voters must assess and articulate their preferences under external constraints; and (2) city government provides limited visibility into the voting process~\cite{PBChicago-PeoplesBudget,PBChicago-rulebook,PBChicago-TIF,Weber2013-civics-PBChicago}.

We study how interactive visualization can augment the voting process---\eg{} by using graphical elicitation techniques to enable voters to draw their budget preferences (\figref{fig:interface}), and by using dashboards to provide transparency around results of the vote (\figref{fig:map}, \figref{fig:strip-plot}).
Our work builds on the idea of using visualizations as input devices~\cite{huron21DataInput}, enabling users to externalize their thinking in ways that can benefit recall of information, understanding of complex processes, and the personalization of displays~\cite{hohman2020communicating,Kim2017-gap,Mahajan2022-vibe,Nguyen2018-BeliefDrivenDJ}. 
This represents an alternative ``semiotic regime'' for visualization~\cite{Engebretsen2020-vis-in-society,Offenhuber2023-autographic}, one in which end-users of a visualization are active co-creators rather than passive receivers of information.
We suggest that these properties make graphical elicitation well aligned with PB's goals of engagement and empowerment.

In this work, we encounter a natural tension between the design goals of (1) affording PB voters flexibility in expressing their preferences and (2) placing those preferences in dialogue with the normative framework of voting.
Failure to address this tension might result in a PB process where votes are easily counted but do not actually reflect public priorities.
We try to guard against potential pitfalls of technological solutionism~\cite{Rosner2018-CriticalFabulations}---\eg{} that software might bias results of PB by skewing participation toward the advantaged or otherwise reifying power dynamics that limit the freedom and agency of voters~\cite{Stortone2015-hybridPB,Weber2013-civics-PBChicago,Nelimarkka2019-democraticDM-review,Harding2015-civicHCI-trust,Corbett2018-engagement,Erete2017-empowered-participation}. 
By investigating how visualization can contribute to transparency and accountability in PB, by studying how graphical elicitation can faithfully reflect people's preferences,  and by exploring the challenges of municipal governance and civic engagement~\cite{Harding2015-civicHCI-trust,Corbett2018-engagement}, we attempt to find ways that visualization can promote more fairness and freedom of expression in digital democracy applications.

To this end, we contribute a formative interview study (N=13) characterizing the perspectives of policy workers and academics with expertise in PB, urban planning, and civic HCI on the role of visualization software in PB.
We use a design probe~\cite{Hutchinson2003-DesignProbe} to prompt conversations about what challenges and opportunities using interactive visualization for PB poses for governance, and we summarize these perspectives through a qualitative analysis. 
Our probe consists of a web-based version of Chicago’s current PB voting process~\cite{PBChicago-rulebook,PbChicago} extended with graphical elicitation methods mediating the vote and a dashboard enabling residents to monitor results. 
We find that visualization can facilitate (1)~voting interfaces that give city residents greater freedom to express their preferences, and (2)~deliberation activities and tailored data storytelling that expand civic capacity through voter education and outreach. 
We argue that realizing these benefits will require the visualization community to view PB (and democracy) not as just another application domain but as part of the fabric of civic life that deserves deep, sustained efforts to study and cooperatively design political processes.
See \secref{sec:supp} for links to our probe and supplemental materials.


\textbf{Positionality statement.} We are visualization researchers at the University of Chicago, a wealthy institution situated in the historically disinvested South Side of Chicago. 
Although we undertake this work to explore how our research might benefit the community around us, we also recognize that we operate in a context of deep structural inequities, and that even careful study of PB interfaces for Chicago risks exercising power in unintended ways.
As such, we adopt a need-finding stance and avoid prescriptions about things like the voting behavior of residents.

%% file: 2_related.tex
\section{Background}
\label{sec:background}
Our study explores how software for participatory budgeting (PB) in Chicago might be augmented by interactive visualization, including the use of graphical elicitation techniques to measure the preferences of city residents.
We contextualize our contributions in relation to Chicago's existing PB process, tools for digital democracy, previous work on graphical elicitation, and perspectives from utility theory.

\begin{figure}[t]
    \centering
    \includegraphics[alt={A mock ballot resembling Chicago's current PB vote elicitation. The ballot is a checkbox list with a description, estimated cost, minimum required cost, and a representative image for each of the following projects: picnic tables, curb cuts, street resurfacing, bike lanes, steet lights, street murals.},width=\linewidth]{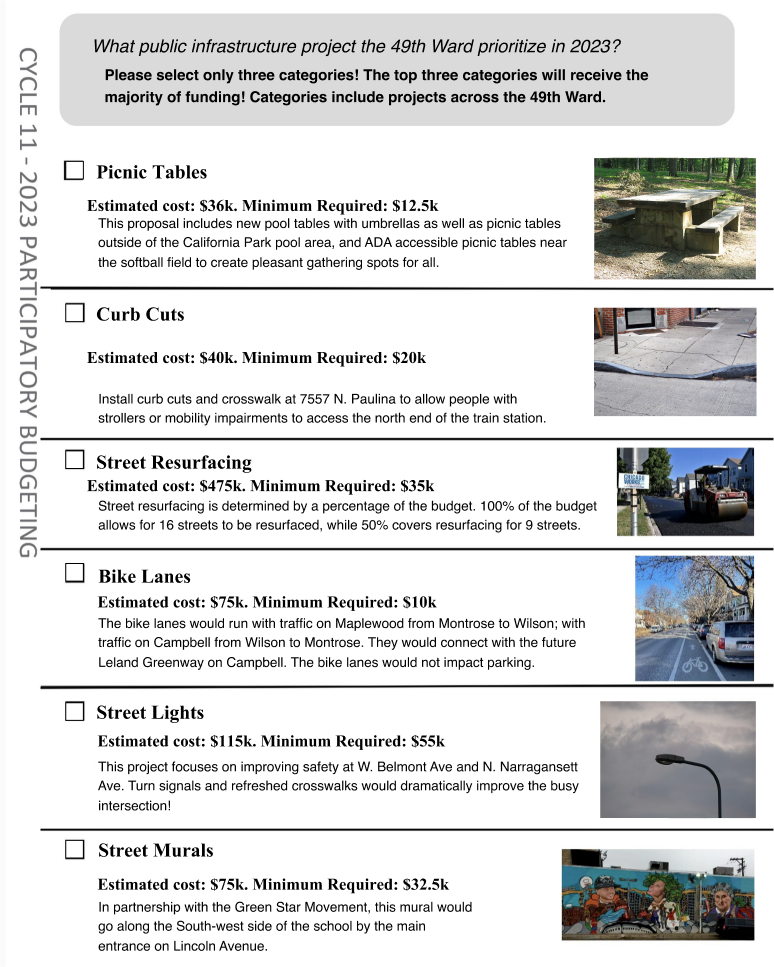}
    \caption{Current PB ballots tend to be a static paper-like form which include cost estimates, project descriptions, and sample images. Here we have adjusted an real ballot to describe the projects used in our probe.}
    \label{fig:ballot}
    \vspace{-6mm}
\end{figure}

\subsection{Participatory Budgeting in Chicago}
\label{sec:pb-chi}
PB is a process of direct democracy, often practiced at the municipal level, where government asks residents to weigh in on how discretionary funds are spent.
This practice originated in Brazil in 1989 as an antipoverty measure, which helped to empower the public and dramatically reduce childhood mortality~\cite{Wampler2014-WaPost}.
Since then it has been adopted in cities around the world in the hope of engaging residents about their priorities and improving equity in government spending~\cite{Palacin2023-PB-configurations,StanfordPB}.

Chicago was the first city in the United States to adopt PB, where since 2013 it has been used by a subset of the city's 50 \textit{wards} (i.e., city council districts) to distribute an annual pot of up to $\$1.5M$ USD called \textit{``menu money''}~\cite{Weber2013-civics-PBChicago,PBChicago-TIF}.
The \textit{alderperson} (i.e., city councilperson) in each ward may only spend menu money on certain types of public infrastructure, such as road repairs, parks, and school improvements. 
See the Chicagoist~\cite{change19Chicagoist} for an overview of Chicago's political structure.

Early in Chicago's PB process, residents propose projects for consideration, which are then vetted for legality, feasibility, and cost by committees of community members and ward staff~\cite{PBChicago-rulebook,PBChicago-PeoplesBudget}.
Only a handful of these proposals end up on a ballot where residents are asked to  a vote on a subset of projects to fund.
The alderperson takes these votes into consideration but ultimately has prerogative over which projects are implemented.
Participation rates in Chicago's PB process are low ($0.23-2.28\%$ per ward), however, it engages 
\revision{a greater proportion of}
residents from marginalized groups than local elections, especially when PB is paired with targeted outreach efforts and mobile voting options~\cite{Weber2013-civics-PBChicago,PBChicago-PeoplesBudget}. 
Our work builds on extensive efforts to implement PB in Chicago~\cite{PbChicago,PBChicago-PeoplesBudget,PBChicago-TIF}.
We investigate how interactive visualization can augment the PB process, in particular exploring design patterns that might make the voting process more engaging and transparent.
\looseness=-1

\subsection{Software for Participatory Budgeting}
\label{sec:pb-tools}
There are many platforms and toolkits designed to support PB (e.g.,~\cite{Consul,Decidim,StanfordPB,PbProject,PbChicago,BiPart-com,OpenNorth-com,Participare-com,CitizensBudget-com}). 
Many of these are open-source projects maintained by academics or non-profit organizations.
Software for PB tends to be feature-rich and highly configurable to meet the needs of any adopting municipality~\cite{Palacin2023-PB-configurations}.
In this work, we emphasize four stages of the PB process that are most commonly instantiated in software~\cite{Palacin2023-PB-configurations,Stortone2015-hybridPB}:
\begin{enumerate}[noitemsep]
    \item \textbf{Idea collection:} An open call for residents to propose and publicly deliberate about potential projects.
    \item \textbf{Proposal assessment:} Multiple processes by which proposals are vetted by community members, policy workers, and domain experts for inclusion on the ballot.
    \item \textbf{Voting:} The process by which residents formally indicate their preferences among selected projects.
    \item \textbf{Results:} An announcement of the which projects will be implemented, chosen based on the vote and aldermanic prerogative.
\end{enumerate}

These stages are typically implemented via standard web UI components, such as forum threads for deliberation and input forms for voting.
In our probe, we investigate uses of visualization for voting and results in particular, since these were use cases where we felt most confident suggesting visualization-based augmentations to Chicago's PB process. 

Although a broad review of tools for civic participation, governance, and deliberative decision-making is out of scope for our work (see Nelimarkka~\cite{Nelimarkka2019-democraticDM-review} for such a review), it is important to acknowledge the many contributions of human-computer interaction in this space. 
In particular, we highlight previous research systems that incorporate visualization.
For example, Viewpoint~\cite{Taylor2012-Viewpoint}, CommunityCrit~\cite{Mahyar2018-CommunityCrit}, CommunityPulse~\cite{Jasim2021-CommunityPulse}, MyPosition~\cite{Valkanova2014-MyPosition}, ConsiderIt~\cite{Kriplean2012-ConsiderIt}, Procid~\cite{zilouchianmoghaddam2015-Procid}, and Pol.is~\cite{Polis} each use visualization to support opinion sharing and public deliberation.
Authors of these systems found affordances of visualization for driving community engagement and discussion of disagreements.
Other tools like Factful~\cite{Kim2015-Factful} and BudgetMap~\cite{Kim2016-BudgetMap} make government budget data more accessible through interactive tagging and visual exploration, suggesting opportunities to broaden access to public data generated by processes like PB (e.g., see Ward Wise~\cite{Wardwise}).
We build on prior work by thoughtfully examining the role of these kinds of interactive visualizations in PB, rather than by exploring a broad space of visualization designs that might be used in PB.

\subsection{Graphical Elicitation}
\label{sec:elicitation}
We use graphical elicitation as a voting interface, enabling PB voters to draw their desired allocation of funding across proposed projects on an interactive bar chart.
This style of graphical elicitation technique was developed primarily to elicit beliefs~\cite{goldstein2014,distBuilder} and has been used in recent evaluations of data interpretation behavior (e.g., \cite{Kim2017-gap,hullman2018-imagining-replications,Kim2019-bayes,Rogha2024-elicitation,Karduni2021-bayes,Koonchanok2021-DataProphecy,Koonchanok2023-elicitation,Mahajan2022-vibe}).
Data journalists (e.g., at The New York Times~\cite{aisch2015-youDrawIt,buchanan2017-youDrawIt,katz2017-youDrawIt}) adapted belief elicitation into a \textit{predict-and-reveal} pattern, where the audience is first asked to draw what they think data will look like and then the true data are revealed on the same chart.
Kim \etal{}~\cite{Kim2017-gap} found that this design pattern improves memorability of data through active engagement.
In designing our probe, we selected graphical elicitation as a design affordance in order to attempt to draw out this behavior.

Relatively little work to date examines graphical elicitation of values, rather than beliefs---i.e., how much value people place on certain outcomes, rather than how likely they think those outcomes are.
One exception is a recent study by Verma \etal{}~\cite{Verma2023-vis-fair-allocation}, which asks participants to allocate humanitarian aid using an interactive visualization to assign dollars to relief efforts.
They find visual allocations unfair, in the sense that not every impacted individual received an equal share of funding from participants, however, leading theories of human decision-making~\cite{Kahneman1979-prospect} imply a more nuanced definition of fairness based on the fact that the marginal value of a dollar depends on a person's circumstance.
Prior work on ``visualizations as data input''~\cite{huron21DataInput} emphasizes possible uses of visualization for polling and structured reasoning, and similarly, work on ``autographic visualization''~\cite{Offenhuber2023-autographic} advances the idea of using visualization to ``actively construct evidence.''
In this work, we ask policy workers and academics to reflect on the kinds of interactive visualization design patterns that can support voters in weighing trade-offs in priorities during the allocation of municipal funding.

\subsection{Utility Theory}
\label{sec:utility}
Our investigation places graphical elicitation in dialogue with ideas about utility elicitation from behavioral economics~\cite{Elicitation2018}. 
Utility elicitation procedures prioritize measuring preferences in ways that conform to von Neumann–Morgenstern's axioms~\cite{vonNeumann1944}, which form the foundation for theories of rational choice. 
The resulting procedures are numerous and tend to disambiguate formal preferences by asking participants to make repeated forced choices among alternatives, to rank outcomes, or otherwise to assign dollar values to hypothetical events~\cite{Elicitation2018,Riabacke2012-SOTA-weight-elicitation}.
We posit that repeated judgments would not be appropriate for a quick voting procedure, so we opt to explore rank- and dollar-allocation-based elicitation procedures in our design probe.

Utility theory plays an important role in elections more broadly---e.g., informing both formal voting methods~\cite{Goel2019-knapsack,gelauff2024-rank} and a theory called the ``calculus of voting'' which gives a rational agent framing of decisions about whether or not to vote~\cite{Riker1968-calculus}. 
For our work, an important question is how to reconcile (1) interfaces that give PB voters more freedom to express their priorities with (2) voting methods that require an unambiguous, normative interpretation of preferences. 
We analogize this to the debate about the value of utility theory in medical decision-making~\cite{Cohen1996-utilityMedicalDM,Nease1996-utilityViolations}, where the preferences of doctors and patients may not strictly follow assumptions about rationality~\cite{vonNeumann1944,savage1972foundations}. 
\revision{We draw inspiration from Nease's argument}~\cite{Nease1996-utilityViolations}
\revision{that utility theory can provide a helpful reference for decision-making processes without being ``purely normative'' in the sense of prescribing a course of action from elicited preferences.}
We treat voting interfaces as tools for thinking rather than formal decision procedures.
\revision{As such, \textbf{we do not assume that PB votes reflect strictly ``rational'' preferences}, nor that they need to in order to provide a valuable source of information for decision-making.}
In this study, we investigate what it means for graphical elicitation 
to achieve valid preference measurements in the context of PB, 
\revision{where faithful description of preferences may require relaxation of theoretical axioms.}


%% file: 3_probe.tex
\begin{figure*}
    \centering
\includegraphics[alt={Screenshopts of three stages of elicitation in our probe. (A) Order elicitation: Users start by sorting a collection of projects based on their values, such as funding bike lanes being more important to them than funding steet lights; pictured is our drag-and-drop sort interface. (B) Allocate preferences: Next users are presented with a bar chart, where they can click and drag bars to allocate a million dollars across each of the proposed projects; pictured is an interactive bar chart showing dollars allocated per project. (C) Check allocations: Finally, users see the amount required for each project and are given another opportunity to adjust their preferences; pictured is the allocation bar chart with estimated costs superimposed on the user's previous allocations.}, width=\linewidth]{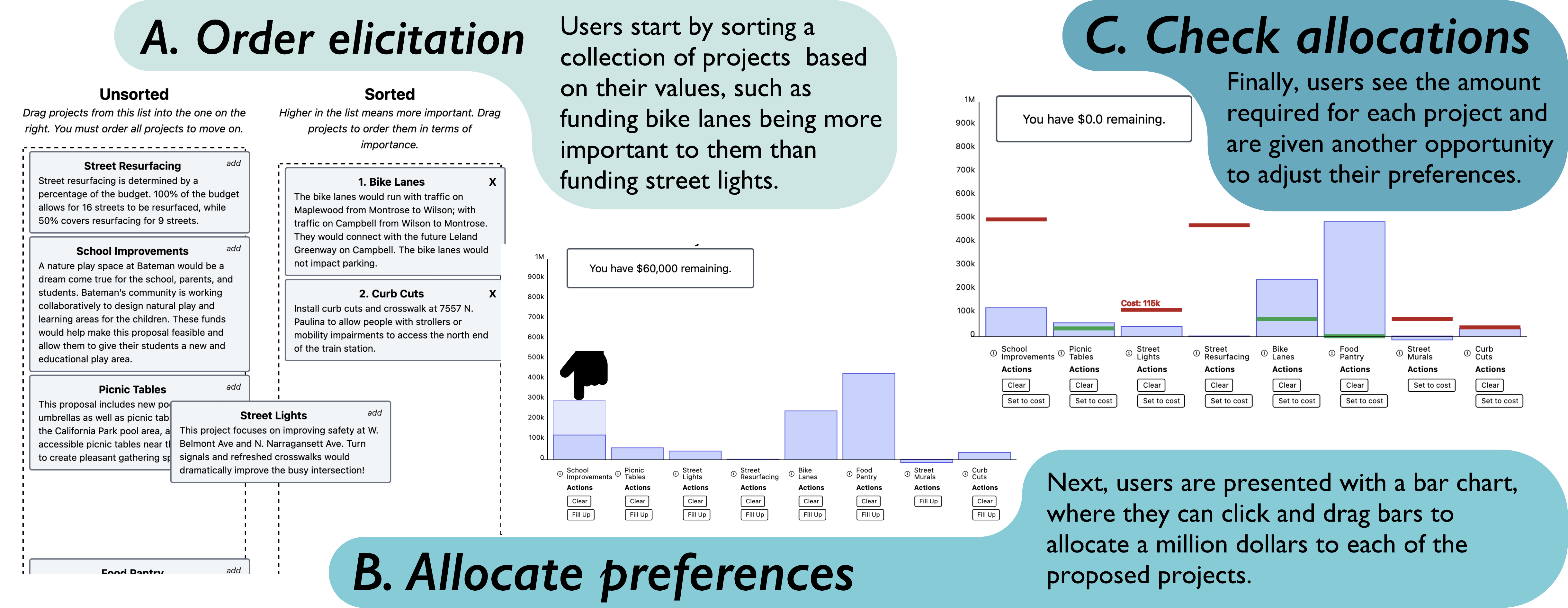} 
    \caption{In our design probe, participants weigh their preferences about projects on a hypothetical PB ballot in three stages.}
    \label{fig:interface}
    \vspace{-4mm}
\end{figure*}

\section{Design Probe}
\label{sec:probe}
We developed an experimental PB process incorporating interactive visualization (\figref{fig:interface}, \figref{fig:map}, \figref{fig:strip-plot}) as design probe~\cite{Hutchinson2003-DesignProbe} to prompt conversations with PB experts.
The probe focused on how 
visualization might 
augment \textit{voting and reporting of results in PB} (see \secref{sec:background}).
We focus on these stages of 
PB
because (1) they are common to most applications of PB~\cite{Palacin2023-PB-configurations}, and (2) we felt more confident about how visualization could helpfully augment 
them versus stages like project ideation and approval.
See \secref{sec:supp} for 
our probe and codebase.

\subsection{Probe Walkthrough}
Imagine a resident of Chicago's 49th Ward, \textit{Karla}, who has been asked to vote on PB projects proposed and vetted by members of her community and the 49th Ward Alderperson's staff.
Like other 49th Ward residents, she receives an email from the ward office asking her to participate in PB.
The 49th Ward is a demographically heterogeneous, majority non-white voting district on Chicago's far north side.
Karla opens the link 
\revision{to}
an \textit{\textbf{overview}} of the PB process. 
It explains what PB is, how projects were selected this cycle, who is eligible to vote, how votes are counted, and how to find updates on the PB process after voting.

Karla 
\revision{begins voting by entering}
the \textit{\textbf{sort}} view, which asks her to order projects based on their importance to her.
The sorting interface uses a drag-and-drop interaction to facilitate quick prioritization of projects with the ability to easily revise (\figref{fig:interface}A).
Once she is happy with her sort order, Karla moves to the next stage of preference elicitation
\revision{where she}
is asked to \textit{\textbf{allocate}} dollars to each project.
The allocation interface provides her with a bar chart to draw on (\figref{fig:interface}B),  such that she can use a drag interaction to directly manipulate the portion of a $\$1M$ budget that she wants to give to each project.
Project are arrayed along the x-axis according to Karla's sort order from the previous page, each initially set to $\$0$.
Buttons beneath each bar enable Karla to ``clear'' the bar by resetting it to $\$0$, or ``fill up'' the bar by allocating the remaining funds in the budget.
A box at the top of the chart reminds her how much of her total $\$1M$ budget remains.
Karla adjusts the bars until they shows her desired distribution of funding across projects.
The next page asks Karla to \textit{\textbf{check allocations}}.
She views the same bar chart as on the previous page, now with the estimated cost of each project superimposed as rule markings 
\revision{that are color coded based on whether the cost of a project is met}
(\figref{fig:interface}C).
Following a prompt, 
\revision{Karla revises}
her allocations now that cost information has been revealed.

An optional \textit{\textbf{demographics}} survey on the next page
asks about race, age, income, and education 
to inform on 
the fairness of the PB vote.
\looseness=-1

The final page provides a lightweight \textit{\textbf{dashboard}} where Karla is invited to explore data about the PB vote so far.
The dashboard contains two primary views: (1) a heatmap (\figref{fig:map}) showing the demographics of PB voters in each ward compared to the census; and (2) a strip plot (\figref{fig:strip-plot}) showing the dollar allocations of individual voters in each ward.
This interface uses drop-down menus, toggle buttons, and an interactive map to enable selection of wards to compare (\ie{} 29th, 35th, 36th, 49th), demographics to view (\ie{} race, age, income, education), and voting statistics (\ie{} counts, percentages).
Karla spends some time inspecting whether the PB vote is representative and the allocations of other voters, which prompts her to reflect on the fairness of the PB process and how her own priorities compare to those of others.

This scenario reflects how we expect residents might interact with a tool like our probe, not how Karla would vote in Chicago's current PB process, which involves an approval vote on a ballot (\figref{fig:ballot}). Ward staff usually follow up with voters through more conventional outreach that does not necessarily involve sharing data about the vote~\cite{PBChicago-rulebook}. 

\subsection{Design Choices \& Rationales}
Our design probe was not intended to propose a fully-fledged PB process or to replace any existing practices, but instead to prompt reflection of PB experts on the use of interactive visualization. 
Our design targeted uses of visualization for reasoning about budget trade-offs and considering the preferences of other voters.
We incorporated interactive visualization in two ways: (1) graphical elicitation of voter preferences; and (2) a dashboard displaying results of the PB vote.

\textbf{Graphical elicitation of the PB vote.}
Rather than a traditional approval vote to fund individual proposals (e.g., the City of Chicago's current PB ballots, \figref{fig:ballot}), we adopted a design pattern where participants were asked about their preferences in three stages: (1) sorting projects, (2) allocating dollars without seeing cost information, and (3) adjusting allocations after seeing project cost estimates.
This voting procedure incorporated a predict-and-real pattern for project costs similar to They-Draw-It!~\cite{Nguyen2018-BeliefDrivenDJ}, which has been found to improve recall and engagement with visualized data~\cite{Kim2017-gap,hohman2020communicating}.
Inspired by the Cardinal Rank Ordering of Criteria (CROC) method for utility elicitation~\cite{Riabacke2009-CROC}, we asked participants to first sort projects and then allocate dollars to them as a way of gradually refining their sense of preference.
In our interviews, this design pattern helped to prompt reflection on the nature preferences and what is actually measured by PB votes.

\textbf{Dashboard of voting results.}
For each ward participating in a hypothetical PB cycle, we show demographic information about who voted and the distribution of voters' allocations.
Inspired by uses of interactive visualization in visual analytics~\cite{Roberts2007-multipleViews} and election coverage by data journalists~\cite{bostock2014-election,aisch2016-election}, we built our dashboard to enable participants to explore the results of PB vote. 
We expected that visualization would be helpful for surfacing potential imbalances of voter representation in comparison to the census, as well as disagreements in funding allocations.
In our interviews, the dashboard helped to prompt reflection on issues of transparency, accountability, and fairness in PB and how (if at all) visualization plays a role in addressing these. 

\begin{figure}
    \centering
    \includegraphics[alt={A contingency table and heatmap showing demographics of a simulated PB vote. The table crosses race (rows) with population and rate of PB participation per ward (columns). The table shows population and participation for two selected wards to facilitate comparisons of the representativeness of PB voting. Controls allow the user to select other demographic variables to show across rows, and to toggle between seeing data as counts vs percentages. Cells are shaded based on the data they contain, and a quantized colorscale legend is provided.}, width=\linewidth]{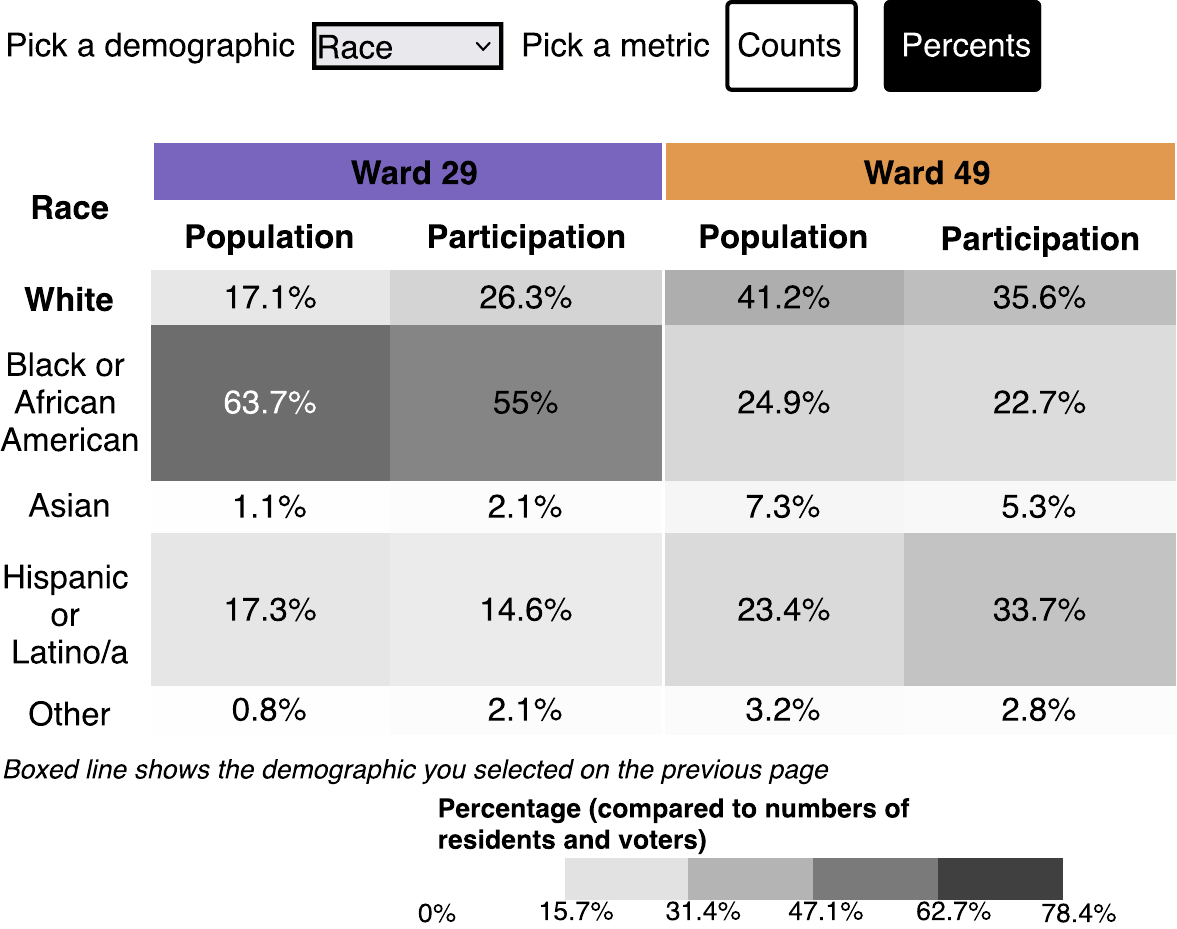}
    \vspace{-1em}
    \caption{On the final page of our probe, participants viewed a dashboard of simulated voting results. This dashboard also included a clickable map (not shown) to select which wards to compare. }
    \label{fig:map}
    \vspace{-2em}
\end{figure}

\begin{figure}
    \centering
    \includegraphics[alt={A strip plot showing simulated allocations to projects about arts and culture, biking and transport, libraries and schools, parks and environment, and streets and sidewalks. Participants in our study were shown similar plots for each ward. Simulated allocations are highest and have the largest variance for libraries and schools, and parks and environment.}, width=0.8\linewidth]{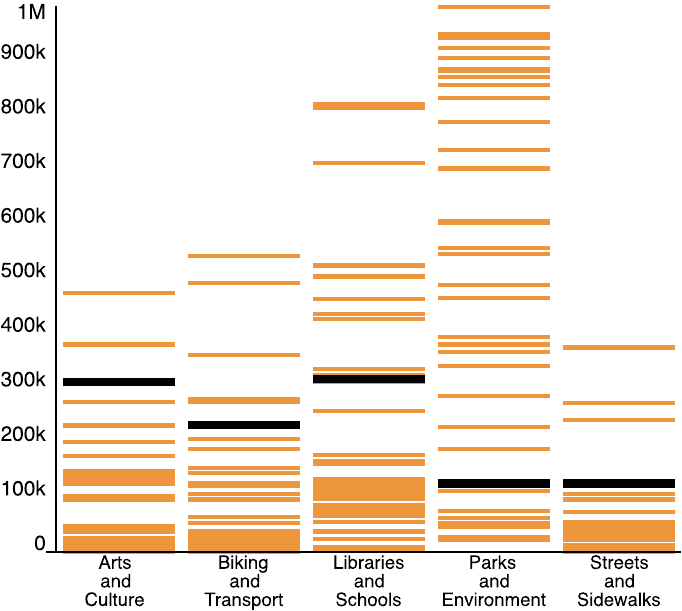}
    \caption{In addition to \figref{fig:map}'s heatmap, our dashboard included a strip plot for each selected ward showing how participants' allocation of funds compared to that of (simulated) others.}
    \label{fig:strip-plot}
    \vspace{-1.2em}
\end{figure}

\textbf{Alternative visualization designs.}
Our team considered other designs, but our goal was to probe whether using visualizations for preference elicitation and transparency might be helpful for PB, not to evaluate design variations.
For example, we thought of using pie charts to elicit votes as proportions of the budget because they naturally convey part-to-whole, but we thought this would lead to more difficult labeling and interaction design.
We opted to show a heatmap of demographic comparisons (\figref{fig:map}) rather than mirrored bar charts, and a strip plot of allocations (\figref{fig:strip-plot}) rather than a density plot, because we wanted to avoid visualization design choices such as mirrored axes and smoothing that may require higher graphical literacy to decode. 

\subsection{Simulating PB Data}
The data we showed participants in the probe (\ie{} projects, costs, voter demographics, and allocations) was simulated based on previous rounds of PB in Chicago.
PB projects participants were asked to consider and their estimated costs were synthesized from an archive of ballots maintained by the organization PB Chicago~\cite{PbChicago}.
We simulated data on voter demographics based on 2020 census data, curated by PB Chicago to estimate the population demographics of each ward.
Our simulation assumed a rate of participation between 1 and 3\% for each demographic category, consistent with rates of participation in recent runs of PB in Chicago~\cite{Weber2013-civics-PBChicago}. 
We drew hypothetical voters from the census population using a binomial distribution to achieve a realistic degree of sampling error.
To simulate the budget allocations of hypothetical voters, we use a stick-breaking algorithm and assumed that each ward would have an overall propensity toward particular kinds of projects (\eg{} libraries and schools versus streets and sidewalks).
The goal of these simulations was merely to produce realistic-looking data to display.
See \secref{sec:supp} for the scripts used to generate synthetic participation and allocation data.

\subsection{Interface Versions \& Revisions}
During our interviews, participants suggested changes to the design probe.
We implemented these incremental changes between interviews when we felt that they would improve the usability of our probe without fundamentally changing it.
These revisions clustered into three version updates.
The first update entailed \textit{\textbf{simplifying the color scale of the demographic heatmap}}, which initially used separate fill colors to differentiate the census population versus PB vote participation.
We revised this to one gray color scale (\figref{fig:map}) based on the feedback that there were too many colors in the dashboard.
The second change was to \textit{\textbf{revise the interaction design for the sort interface}}, which initially required users to click boxes for each project to add them to their sorted list. 
The redesign involved the addition of an unsorted list where all projects start out (\figref{fig:interface}A), and the ability to drag projects freely between the sorted and unsorted lists.
The third update was to \textit{\textbf{add a granularity setting for the drawing interaction used to allocate dollars}}, which enabled us to control the sensitivity of the drawing interaction. 
The earliest version of the interface allowed allocations to vary with floating point precision, which made it difficult for participants to hit a particular dollar amount.
The revised interface used funding increments of $\$1000$, making it easier for participants to spend exactly $\$1M$ total.


%% file: 4_method.tex
\section{Method}
We set out to answer the question, \textit{How can interactive visualization benefit participatory budgeting (PB)?} with an emphasis on formative need-finding. 
To do this, we recruited experts in PB, technology for civic engagement, and urban planning, and we interviewed them about the role of interactive visualization and software more broadly in PB.
We instigated these conversations using a design probe method~\cite{Hutchinson2003-DesignProbe}, where participants were asked to think-aloud while exploring a lightweight web-based PB procedure augmented with interactive visualizations (see \secref{sec:probe}).
Then, we conducted semi-structured conversational interviews with participants, addressing questions of (1) how to measure public preferences through PB, (2) how to address issues of transparency and accountability in PB, and (3) requirements for usability and adoption of PB software. 
We used qualitative analysis to characterize participants' reactions to the design probe and their reflections on the potential of visualization for PB.

\subsection{Participants}
We recruited 13 participants.
We refer to them by their ids (e.g., \pxx{x}) and highlight their quotes for readability,
\qt{using pale blue from the Chicago flag}. 
These were a mix of academics and policy workers with expertise relevant to our investigation of PB (\autoref{tab:participants}), such as experience implementing PB or researching technology for civic engagement and urban planning.
Our recruiting method was a combination of convenience and snowball sampling.
We screened participants based on the criterion that they have expertise in PB or related topics.

\begin{table}
\caption{Relevant experience of interview participants.}
\label{tab:participants}
\vspace{-2mm}
\begin{tabular}{ |p{2.65in}|p{0.45in}| }
    \hline
    \textbf{Participant background} & \textbf{Ids} \\
    \hline
    Policy workers in Chicago Ward offices responsible for implementing PB & \pxx{01}, \pxx{04}, \pxx{05} \\
    \hline
    Policy workers at non-profits or academic institutes responsible for implementing PB & \pxx{02}, \pxx{03} \\
    \hline
    Academics whose research involves implementing PB & \pxx{09}, \pxx{10} \\
    \hline
    Academics with expertise in HCI and digital civics & \pxx{06}, \pxx{07}, \pxx{12}, \pxx{13} \\
    \hline
    Academics with expertise in urban planning & \pxx{08}, \pxx{11} \\
    \hline
\end{tabular}
\vspace{-1.5em}
\end{table}

\subsection{Procedure} 
Our interviews were split into two blocks.
In the first block, we asked participants to \textit{\textbf{think-aloud}} while using with the design probe:
\begin{quote}
\vspace{-2mm}
    Imagine that you are a resident of the 49th Ward giving input on funding for projects suggested and vetted by members of your community. As you use the application, please say what you are thinking out loud. We are particularly interested in your impressions of the way we use interactive visualizations, how tools like this might be used in actual governance, and any questions or concerns that come to mind for you when using the tool.
\end{quote}
\vspace{-2mm}
They stepped through a series of web pages in our probe, first sorting and allocating money to PB projects and then viewing the results dashboard.
The interviewer refrained from interjecting other than to answer direct questions or to remind the participant to think-aloud.

In the second block, we conducted a \textit{\textbf{semi-structured conversational interview}} around the following discussion topics:
\vspace{0.25em}
\begin{enumerate}[noitemsep]
    \item What we measure in PB, and how voting procedures change this.
    \item Transparency, accountability, fairness, and trust in PB.
    \item Adoption and usability of PB software like our design probe.
\end{enumerate}
We asked each participant the same questions following an interview guide (see ~\secref{sec:supp}).
Participants reflected on our design probe and their experiences with PB, sharing their expert perspectives.

Interviews were conducted and recorded using Zoom meeting software. 
Audio recordings were transcribed for analysis.
Additionally, our design probe logged participants' budget allocations for analysis.

\subsection{Qualitative Analysis}
We conducted a qualitative analysis of interview transcripts.
Starting with open coding, two authors reviewed each interview recording and identified episodes of interest that were relevant to our research questions, documenting quotes and taking notes to preserve context.
We met frequently to ensure consistency in our open coding processes.

After the identification and description of episodes of interest through open coding, we discussed patterns that emerged across interviews, resulting in inductive codes for reasoning about trade-offs, transparency, access issues, and affordances of online PB.

%% file: 5_results.tex
\section{Results}
Our interviews elicited discussions about the role of interactive visualization and software more broadly in PB.
We summarize these discussions for the visualization community focusing on how interfaces frame preferences, the fraught relationship between transparency and trust, access concerns, and possibilities for future online PB tools. 

\synthbox{We bookend topics of interest with \textit{\textbf{synthesis boxes}} where we highlight findings, draw connections to related work, and point out opportunities for future work.}

\subsection{Reasoning about Value Trade-offs}
\label{sec:value-per-dollar}
The core task facing PB voters is to decide how to fund projects based on their judgment of costs and benefits.
Our design probe prompted participants to reflect on the ways these value judgments might be influenced by the design of the voting interface and the provision of information about proposed projects.

\textbf{Interactive visualization is a promising tool for supporting reasoning about trade-offs in priorities.}
Participants (\pxx{06}, \pxx{11}, \pxx{12}) said that the immediate visual feedback provided by the interface facilitates a form of perspective taking where PB voters are forced to confront the practical constraints faced by policy makers: \qt{If I raise this bar up, I can't raise other bars up. It makes very concrete the reality of having a pie that can only be sliced so many ways.} (\pxx{08}).
\pxx{11} and \pxx{12} appreciated how our graphical elicitation methods provide reactive validation that does the math for voters (\eg{} dollars remaining when allocating) and prevents response errors.
Other reasons for preferring graphical elicitation to a conventional ballot were that approval voting is not expressive enough to capture voters' preferences (\pxx{12}, \pxx{13}), whereas our method of allocating dollars across projects \qt{honors people's preferences} (\pxx{02}) more than approval voting (\pxx{02}, \pxx{12}, \pxx{13}).

In particular drawing a distribution of dollars across projects enables \textit{partial allocations}, where PB voters give more or less than the estimated cost to a proposed project.
\pxx{08} and \pxx{12} described how partial allocation carries important information---e.g., under-allocation can signal desire for cheaper projects, and over-allocation can signal desire for more investment than was proposed.
This provides greater flexibility in voting on mundane-seeming and expensive projects (e.g., road repair) that might never get selected for full funding under an approval voting model (\pxx{01}).
However, some (\pxx{02}, \pxx{07}, \pxx{09}) expressed concern that enabling allocation without constraint might give voters miscalibrated expectations about how funds can be spent.
For example, during the think-aloud two participants initially misinterpreted the estimated project costs as minimum (\pxx{06}) or maximum spending limits (\pxx{08}).

\synthbox{Our probe offered participants maximum flexibility to articulate preferences within the constraints of the PB process. Although some of these constraints are compatible with utility theory (\eg{} allocating dollars from a limited budget),  others are not (\eg{} considering only pre-approved projects), suggesting that our analogy between utility elicitation and PB voting (see \secref{sec:utility}) is imperfect. Nonetheless, our tool produces more information-rich preference measurements than traditional approval voting, and participants (\pxx{01}, \pxx{02}, \pxx{09}) discussed how allocations might be compatible with formal vote counting methods such as knapsack voting, ranked choice voting, or equal shares voting~\cite{Goel2019-knapsack,gelauff2024-rank}.}


\begin{figure}[t]
    \centering
    \includegraphics[alt={Spaghetti plots per participant showing how the rank order of projects changed across the three stages of elicitation: sort (no cost information), allocate funds (users specify amounts), check allocations (project costs revealed). Ranked projects were bike lanes, curb cuts, food pantry, school improvements, picnic tables, street lights, street murals, and street resurfacing. Spaghetti plots give the overall impression that priorities tend to reorder across stages, and that priorities are different across participants.}, width=\linewidth]{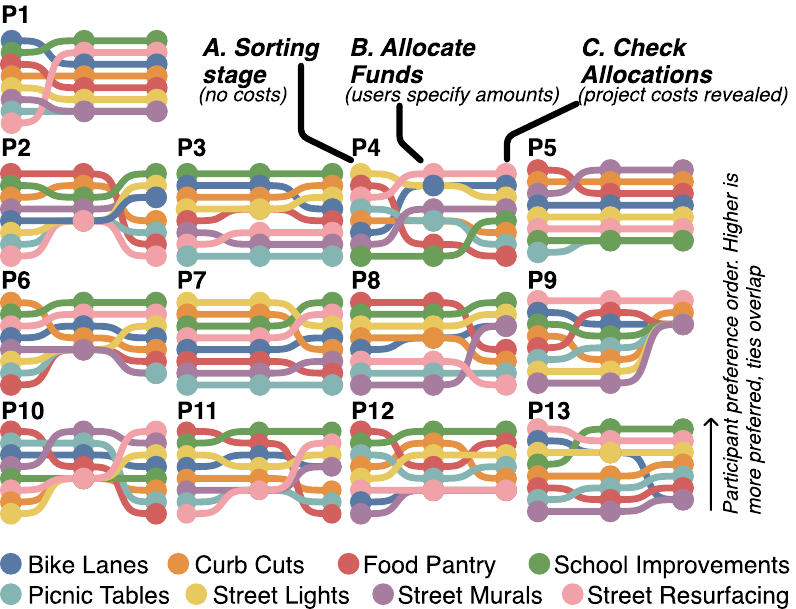}
    \caption{Participants typically changed their preferences in light of the new information revealed at each stage of the design probe. 
    }
    
    \label{fig:spaget}
    \vspace{-6mm}
\end{figure}

\textbf{Does framing preferences in terms of dollars make people's sense of value more real, or does it distract from their priorities?}
Participants (\pxx{09}, \pxx{11}, \pxx{12}, \pxx{13}) suggested PB interfaces should help voters separate considerations of expected cost from a project's relative importance or a community's need.
Although all participants said they wanted our probe to reveal project costs earlier in the voting process, they disagreed on the role that cost should play in informing PB voting.

\pxx{02} and \pxx{09} emphasized the importance of understanding costs, suggesting that dollar allocations would otherwise be arbitrary. 
\pxx{01} was more skeptical that interfaces can make cost information comprehensible to PB voters:
\qt{I've done PB for 4 years, working on a 5th... I think that we should be asking people to state what their priorities are, like, `What do you want to see?' but not how much money they want to allocate, because at the end of the day even I struggle to see what this money means.}.
\pxx{01}, \pxx{08}, and \pxx{09} noted that reasoning about the value of projects in dollars is difficult even for policy workers, leading \pxx{01} and \pxx{13} to suggest that it might be sufficient if the elicitation procedure stopped at ranking projects in order of importance.

Elicited preferences became more about proposed spending and less about about perceived need or importance when participants considered units of dollars (\pxx{07}, \pxx{09}, \pxx{11}) or cost estimates (\pxx{01}, \pxx{08}, \pxx{10}).
Interaction logs from our interviews (\figref{fig:spaget}) were consistent with the idea that participants' thinking pivoted both when dollars were introduced and when cost information was revealed.
For example, \qt{These horizontal lines very heavily influenced my decision making in terms of how much to allocate, because I'm assuming that once this horizontal line is met, everything is great, and this project can move forward.} (\pxx{08})
In particular, the ``fill up'' and ``set to cost'' buttons at the bottom of the allocation bar chart (\figref{fig:interface}B-C) induced a cognitive shortcut by inviting people to accept the estimated cost (\pxx{08}, \pxx{11}).

Some participants (\pxx{08}, \pxx{12}, \pxx{13}) raised the issue of how to handle uncertainty in costs and wanted the interface to distinguish between projects that were all-or-nothing versus projects that could be funded to varying degrees to achieve lesser or greater impact.
They suggested presenting costs as ranges or as thresholds representing feasible funding levels, suggesting this might lead to better-informed partial allocations.
However, \pxx{09} was concerned that ranges or thresholds would add too much cognitive load to the voting procedure.

\synthbox{We chose to emphasize project costs based on previous PB ballots~\cite{PbChicago} (\figref{fig:ballot}), and our probe elicits dollar allocations following work in behavioral economics~\cite{Elicitation2018}. 
Our interviews suggest future work on uncertainty visualizations for projected municipal spending (e.g., ranges, thresholds, or semiotic cues like blur or sketchiness), which could be used to inform PB voting but might also play an educational role in calibrating residents' expectations.}


\textbf{The role of project context in value judgments.}
All participants wanted more visibility into the context of proposed projects in order to more easily assess project importance and community need.  
For example, \pxx{09}, \pxx{10}, and \pxx{12} wanted before-and-after pictures to make the stakes of proposed projects clearer, likely in reference to prior work~\cite{StanfordPB} which offered that affordance or the current Chicago PB ballot that uses designs from that work (\figref{fig:ballot}). 
Some echoed the desire to know more about the current situation in the ward that a proposed project would address (\pxx{06}, \pxx{07}, \pxx{08}), where in the ward a proposed project would be expected to have an impact (\pxx{02}, \pxx{09}, \pxx{12}), or how many people would be impacted by a project (\pxx{08}, \pxx{11}, \pxx{13}).
In order to make the stakes of proposed projects more real and personal, \pxx{12} suggested that project narratives could be written from the perspective of the residents, rather than policy makers.
However, \pxx{08} raised a concern about electioneering (i.e., illegally attempting to sway voters at a polling location), which problematizes calls to include persuasive contextual information in PB voting interfaces. 

\synthbox{Future work should establish ethical standards and audit processes for how we inform voters about project context in PB ballots. Although Chicago's ballots already include descriptions of and images for each PB project (\figref{fig:ballot}), it's easy to imagine a failure mode where deceptive (\eg{}~\cite{Correll2020}) or affective (\eg{}~\cite{Lan2024-affectiveVis}) visualization design practices could be optimized to persuade~\cite{Bergemann2019-infoDesign,Kamenica2011-BayesPersuasion} voters to take any position.}


\subsection{Transparency and Building Trust}
\label{sec:transparency}
Our interviews revealed that transparency is necessary but not sufficient for building trust.
\revision{In some situations, participants worried that transparency might even promote mistrust.}
We highlight areas of agreement and disagreement between our participants to unpack this tension.

\textbf{Revealing project costs can invite scrutiny.}
Participants bemoaned that sharing project cost estimates can lead to perceptions among voters that cost are exorbitant because of corruption (\pxx{08}, \pxx{10}) or bureaucratic wastefulness (\pxx{12}).
However, \pxx{02} noted how transparency about costs helps to calibrate expectations, which in turn facilitates trust.
Knowing more about how cost estimates are generated could help residents decide not only how to vote but whether cost estimates themselves should make voters mistrustful of government (\pxx{02}, \pxx{08}, \pxx{12}).

\textbf{Transparency about voting results is beneficial.}
Our design probe posed a scenario in which residents learn how their neighbors voted (\figref{fig:strip-plot}) immediately after taking part in the PB vote.
Participants agreed that sharing this data with voters would be helpful, either because residents might want to know what their neighbors voted for (\pxx{01}, \pxx{06}, \pxx{08}) or because seeing the allocations of other residents might prompt creative problem solving (\pxx{02}) or even altruistic voting (\pxx{07}).

\synthbox{Our interviews raised questions about how providing voters context about project costs or preliminary voting results would impact voting behavior and trust in the PB process. Future empirical work is needed to address these questions, which will in turn help to establish ethical standards against electioneering in PB voting interfaces (see \secref{sec:value-per-dollar}).}

\textbf{Transparency about voter demographics is highly contentious.} 
Upon seeing a heatmap of voter demographics in our probe (\figref{fig:map}), participants worried that: (1) the appearance of low turnout or demographic imbalance would invite criticism of PB (\pxx{01}, \pxx{10}); (2) showing who is voting may encourage strategic efforts to herd the vote (\pxx{09}); and (3) asking for demographics in a voting interface might give the impression that votes are not counted anonymously (\pxx{07}, \pxx{09}, \pxx{12}).
For example, \qt{If I say that I'm really young, does it mean that my [vote] does not matter as much as as someone else?} (\pxx{12}).
Participants warned that emphasizing voter demographics in PB stirs up distrust and division:
\qt{When you show demographics and you show education and income, does it reinforce an us-versus-them kind of reaction?} (\pxx{13}).

However, a subset of participants believe that tracking and displaying voter demographic data is essential for PB.
\pxx{08} thought demographic data about his ward could help him make more equitable decisions about what to fund.
\pxx{02} and \pxx{11} argued that transparency about demographics is needed to signal the legitimacy of the PB vote to residents.

\synthbox{Future work involving city residents is needed to find ways of posing questions about demographics more constructively. Visualization typically attempts an epistemological stance of neutrality (\eg{} Haraway's ``god trick''~\cite{haraway2016situated}), which may be insufficiently expressive to meet the situated needs of PB. For example, our probe presented demographics in comparison to the census, tacitly endorsing a single definition for fairness when it might have been better to ask what voters think (\pxx{11}).}


\textbf{Accountability is about showing what projects actually get implemented.} 
\pxx{01}, \pxx{07}, \pxx{10}, and \pxx{12} mentioned this was a major barrier to building trust.
Transparency about what happens after the vote is a problem in some cities, which do PB without tracking discrepancies between what voters approve and what ultimately gets done (\pxx{08}, \pxx{11}).
\pxx{02} suggested that a PB interface like our probe could highlight not only discrepancies between winning projects and what gets implemented, but also between projects proposed by the community and those on the ballot. This would reveal bias in proposal selection, in addition to circumventing failures to follow up after the PB vote.

\synthbox{Our probe did not provide strong visual cues that elicited preferences may not match the final outcome of the vote. Rather we provided a text description of the voting process following Chicago's PB ballots~\cite{PbChicago}. Future work should investigate how visualizations of results can use semiotic cues (e.g., sketchiness) in combination with text description to calibrate expectations and facilitate follow-up communications with PB voters (e.g., via email or in-person outreach).}


\subsection{Access to Information and the Vote}
\label{sec:access}
Using software to mediate democratic processes like the PB vote creates possible access barriers for residents, which is a well known issue in digital civics~\cite{Stortone2015-hybridPB,Weber2013-civics-PBChicago,Nelimarkka2019-democraticDM-review,Harding2015-civicHCI-trust,Corbett2018-engagement,Erete2017-empowered-participation}.
If people vote in an interface like our design probe, \qt{It may not be the people that you have a hard time reaching in the first place.} (\pxx{09}).
Specifically, \pxx{07}, \pxx{09}, \pxx{10}, and \pxx{13} speculated that the resulting sample of voters could skew younger, whiter, more educated, and more affluent than the general population.
However, previous PB efforts in Chicago managed to increase participation among marginalized communities by combining in-person outreach with mobile voting software~\cite{Weber2013-civics-PBChicago}.
We unpack the reasons participants gave for their access concerns and possible solutions raised.

\textbf{Is access to the PB vote even a technology problem?}
Our participants seem to disagree on the importance of interface design to PB access.
\pxx{10}, \pxx{12}, and \pxx{13} claimed that the primary access barriers in PB have to do with awareness (e.g., knowing when to get involved~\cite{Erete2017-empowered-participation}) and that the design of voting software doesn't necessarily help to build this awareness.
\pxx{13} summarized this stance:
\qt{Some of those [issues] have nothing to do with the technology piece. But it's really like, are you making it easy for people to participate just given the rhythm of their lives, particularly for low income segments of the population?}
However, part of this broader sociopolitical context is that for many residents, the PB vote is the only time they will engage with local government outside of a general election.
For this reason, \pxx{02} and \pxx{11} suggested that visualizations in a voting interface would create new opportunities to access information about the PB process.
More generally, \pxx{05} and \pxx{09} said that the usability of voting interfaces matters a lot for convincing people to participate, especially for avoiding voter attrition.

\synthbox{Almost all participants agreed that voting software needs to be paired with in-person outreach efforts, a process called ``hybrid PB''~\cite{Stortone2015-hybridPB}, which we revisit in \secref{sec:chal-opp} and \secref{sec:discussion}.}

\textbf{How accessible is the voting interface?}
Our design probe relied primarily on drag interactions to facilitate sorting and drawing elicitation techniques.
While some (\pxx{05}, \pxx{12}, \pxx{13}) found these interactions intuitive, quick, and easy to use, other participants (\pxx{07}, \pxx{08}, \pxx{09}) found these interactions a bit \qt{finicky} (\pxx{07}).
In particular, almost all participants struggled to spend exactly the desired amount without using the ``fill up'' and ``set to cost'' buttons.
Additionally, \pxx{04}, \pxx{06}, and \pxx{11} raised concerns about how this kind of elicitation approach would work for people who use screen readers or other assistive technologies.

\synthbox{Future work should develop analogous preference elicitation interfaces across across graphical, paper, and screen reader media. Critically, empirical testing must validate that these alternative voting methods produce comparable results. We posit that dot-stacking elicitation (\eg{}~\cite{distBuilder}) would be well suited to this purpose because it is as expressive as our probe (\figref{fig:interface}) and translates to bubble-fill and button-press inputs.}

\textbf{What about graphical literacy?}
Participants questioned whether the general public would be able to interpret the visualizations in our probe's dashboard (\figref{fig:map}, \figref{fig:strip-plot}).
\pxx{01}, \pxx{07}, \pxx{08}, and \pxx{10} said it was an information overload, perhaps even for policy makers.
One participant joked that, \qt{This is like a quiz on if I can read standard infographics.} (\pxx{13}).
Specifically, participants found it difficult to interpret:
\begin{itemize}[noitemsep]
    \item The nested layout of the heatmap (\figref{fig:map}) (\pxx{01}, \pxx{09}, \pxx{13}).
    \item The normalization of census and vote counts (\pxx{01}, \pxx{07}).
    \item The use of one mark per voter and bins of project categories in the strip plot (\figref{fig:strip-plot}) (\pxx{01}, \pxx{06}, \pxx{08}, \pxx{11}, \pxx{13}). 
    \item Discoverability of interactive selections (\pxx{02}, \pxx{11}).
\end{itemize}
However, the shortcomings of dashboards for a PB audience go far beyond confusion in decoding design choices.
Participants seemed to reject open exploration as a design goal:
\qt{Don't get me started on dashboards. It's like, Oh, here's the data, tada! So what? ...There is no such thing as raw data, or unbiased data. We always package it. It has an aesthetic, and it's meant for something. So let's just be clear about what is meant for, and also reduce the cognitive burden on the user to know what they're looking at.} (\pxx{11}).
Either in lieu of dashboards or as a guide to learning how to interpret them, \pxx{01}, \pxx{02}, \pxx{09}, \pxx{10}, and \pxx{11} called instead for narrative presentations distilling important patterns and answering critical questions from voting data.

\synthbox{Future research should involve residents and investigate what kinds of narrative visualization design patterns can improve access to data about the PB process. Prior work on visualization onboarding~\cite{stoiber2023visahoi} will be highly relevant to this effort.}


\subsection{Challenges and Opportunities of Online PB}
\label{sec:chal-opp}
With most of the PB vote moving online in some communities, our participants suggested exciting ways that interactive visualization and software more broadly can contribute to the implementation of PB.

\textbf{Visualization-mediated deliberation.}
\pxx{02}, \pxx{06}, \pxx{08}, \pxx{09}, and \pxx{13} suggested that interactive visualization would be useful earlier in the PB process when residents deliberate about what proposals should be considered for the PB ballot.
Similarly, \pxx{01}, \pxx{06}, and \pxx{10} saw the use of graphical elicitation in our probe as more useful for surfacing information about what people prioritize than for assessing whether they agree with spending an exact dollar amount.
Our allocation process was received as an exercise in describing \qt{what I find sensible as an amount for money for these kind of [projects]} (\pxx{12}).
\pxx{02} described how this exercise would be most valuable during the idea collection stage of PB when \qt{the magic of PB} offers a chance to come up with imaginative proposals that honor people's preferences, rather than during the PB vote when policy workers have already assessed feasibility and estimated costs.
\pxx{08} described how participation in this early stage likely raises awareness and buy-in for subsequent PB votes. 

\synthbox{We view this as a call for PB interfaces that are more like public forums, in line with prior work on civic HCI~\cite{Kriplean2012-ConsiderIt,Valkanova2014-MyPosition,Jasim2021-CommunityClick,Jasim2021-CommunityPulse}. 
For example, \pxx{06}, \pxx{08}, and \pxx{09} wanted ways to link public-facing comments to their budget allocations, so residents can share and discuss their rationale.}


\textbf{Avenues for situated voter education.}
\pxx{10}, \pxx{11}, and \pxx{13} interpreted both our uses of interactive visualization (\ie{} for elicitation and data exploration) as having an educational purpose.
For example, the predict-and-reveal design pattern we used during the allocation procedure creates a learning moment from \qt{sticker shock} (\pxx{10}), which prompts voters to reflect on what they believe projects should cost and whether that is reasonable (\pxx{01}, \pxx{09}, \pxx{10}).
\pxx{02} and \pxx{13} noted that this kind of informal learning builds the capacity for people to get involved in the PB process in an ongoing way, not just to vote. 

\synthbox{In \secref{sec:discussion}, we discuss this goal of capacity building, calling for future work to adapt data storytelling techniques (\eg{} relating data to personal narratives~\cite{white2022-covidDeaths}) to ensure that, \qt{people can see their challenges, their neighborhood, the kind of texture of where they live represented in some way,} as \pxx{02} argued plays an important role in building trust.}

\textbf{Provenance for public priorities and spending.}
\pxx{07}, \pxx{10}, \pxx{11}, and \pxx{13} suggested that tracking how residents' priorities change over time would be useful for policy workers.
However, cities struggle to share data about anticipated outcomes of PB proposals (\pxx{09}), demographics of PB voters (\pxx{01}), the geographic distribution of proposed projects (\pxx{02}, \pxx{09}), or even the amount of money spent on particular projects over time (\pxx{04}, \pxx{05}).
Collating such data in a usable format would facilitate feedback loops aimed at making the PB process more equitable (\pxx{07}, \pxx{13}), \eg{} monitoring whether the distribution of funding is geographically equitable (\pxx{02}, \pxx{09}), prioritizing outreach efforts (\pxx{06}, \pxx{09}, \pxx{13}), and knowing when to seek supplemental funding (\pxx{09}).
Historical data can also provide context about the community's need for projects (\pxx{07}, \pxx{10} \pxx{11}, \pxx{13}) and can be used to demonstrate past successes of the PB process to residents (\pxx{11}). 

\synthbox{Although the reasons why cities struggle with data management practices 
are beyond the scope of this work (see Data Feminism~\cite{DIgnazio2020-dataFeminism} for discussion), we call for municipal data centralization initiatives. 
Improving data storage and access standards would open many possibilities for software in governance, which would be a meaningful contribution well within the competencies of computer scientists.
}

%% file: 6_discussion.tex
\section{Discussion}
\label{sec:discussion}
We return now to the question posed in the title of this article: ``What can interactive visualization do for participatory budgeting (PB) in Chicago?''
Our formative interviews with PB experts point to three primary roles that visualization can play in facilitating outreach efforts of municipal government: (1) graphical elicitation as a voting interface, (2) 
\revision{visualization}
as scaffolding for deliberation activities, and (3) data storytelling as a way of building civic awareness and capacity. 
Although we examine PB in Chicago, our findings have broader implications for digital civics and visualization for social good.

\textbf{Graphical elicitation as a voting interface.}
Visualizations provide a powerful metaphor for weighing trade-offs in priorities, especially insofar as they use immediate feedback and response validation to represent constraints of the political process.
Although our probe incorporated complex design patterns such as predict-and-reveal from data journalism~\cite{Nguyen2018-BeliefDrivenDJ} or interactive selections from visual analytics (see \secref{sec:probe}), \pxx{01}, \pxx{05}, \pxx{09}, \pxx{11} and \pxx{12} called for voting interfaces to be more streamlined.
According to the ``calculus of voting''~\cite{Riker1968-calculus}, people will not vote unless the process involves minimal time and effort.
In this sense, using drag interactions to rank priorities (\figref{fig:interface}A) may be sufficient for PB voting, 
\revision{but}
this should be amended with (1) alternative input options (e.g., text boxes or bubble-fill) that translate to screen readers and paper ballots and (2) the option to narrow down larger sets of projects by choosing to rank only a subset.

However, streamlining voting interfaces at the expense of engaging with context such as project costs provides too limited a measure of preferences.
Chicago's current PB process prevents residents from articulating preferences for partial allocations (see \secref{sec:value-per-dollar}). 
This reflects a conceptualization of PB voting---critiqued elsewhere as ``ideal proceduralism''~\cite{Stortone2015-hybridPB}---where the desired result is a forced decision among prescribed options rather than descriptive preference measurement.
Our interviews show that when confronted with fixed costs, \qt{My mentality shifted to, well, let me just click the `fill up' button to meet the green bars, and then it was almost like my priorities just went out the window.} (\pxx{08}).
This suggests a need for voting interfaces that 
\revision{encourage}
residents to freely express their preferences while understanding how elicited preferences map to formal voting methods (e.g.,~\cite{Goel2019-knapsack,gelauff2024-rank}), which policy workers use to make decisions.
\revision{Relatedly, voting interfaces should not attempt to litigate whether preferences are ``irrational'' but instead should explain how stated preferences would be interpreted under a given voting framework and offer residents the chance to revise their responses to ensure that their vote matches their intent.}
We envision that graphical elicitation can empower voters to place their priorities in dialogue with  normative decision processes, delineating their values from constraints of the political process.

\textbf{Scaffolding for deliberation activities.}
Participants who implement PB (see \autoref{tab:participants}) expressed excitement about using 
\revision{visualization}
to scaffold deliberation activities with groups of residents as part of outreach efforts.
Something like our probe's predict-and-reveal for proposal costs (\figref{fig:interface}BC) would help to (1) educate voters by calibrating their expectations about budgets and (2) inform ward offices' approaches to idea collection and proposal assessment.
Tools might prompt ideation about cost-saving measures, ambitious plans to meet community needs, and consensus-building to combat ``individualistic voting''~\cite{Menendez-Blanco-2022-PB-Madrid-biking}.
Deliberation likely plays an important role in disambiguating the relationship between cost and priority, which \pxx{09}, \pxx{11}, \pxx{12}, and \pxx{13} suggest is necessary but seldom an emphasis of PB efforts.

PB platforms, municipalities, and publics tend to focus most on the voting stage of PB, where rates of participation and software support are highest~\cite{Palacin2023-PB-configurations,Stortone2015-hybridPB}.
However, much work in civic HCI more broadly already focuses on digital assemblies for deliberation rather than voting interfaces~\cite{Nelimarkka2019-democraticDM-review}. 
Our research motivates future work using visualization to ``couple digital [assemblies] and analog processes''~\cite{Offenhuber2023-autographic} of situated outreach (e.g., see Viewpoint~\cite{Taylor2012-Viewpoint}, MyPosition~\cite{Valkanova2014-MyPosition}, PosterVote~\cite{Vlachokyriakos2014-PosterVote}, Vote with Your Feet~\cite{Steinberger2014-VoteWithYourFeet}) to facilitate ``hybrid PB''~\cite{Stortone2015-hybridPB} efforts which are necessary to realize PB's promise of community empowerment.

\textbf{Building civic capacity through data storytelling, not dashboards.}
The dashboard of PB voting data in our probe (\figref{fig:map}, \figref{fig:strip-plot}) prompted participants to discuss difficulties that arise around the imperative to be more data-driven.
Limited literacy in visualization, data, and the internet pose challenges for cities, PB organizers, and residents alike, suggesting opportunities for research on measurements and educational interventions to address such literacies (e.g., ~\cite{Boy2014-vislit,ge2023calvi,lee2017-VLAT,Pandey2023-miniVLAT}).

Approaching data sharing from the perspective of visual analytics, it is tempting to provide 
a dashboard \revision{that standardizes} reporting.
However, our interviews made clear that visualization tools for PB need not only to facilitate comparisons of interest (e.g., demographic fairness, ~\figref{fig:map}) but also to provide more narrative framing, exposition, and guidance on how to read charts (e.g.,~\cite{stoiber2023visahoi}).
Such presentations might be personalized based on users' interest and skill (e.g.,~\cite{Schetinger2023-doomDelicious}).
The standardizing approach of dashboards may be poorly suited for PB due to highly heterogeneous municipal data standards (\pxx{02}, \pxx{05}, \pxx{10}) and the messiness of civic engagement~\cite{Corbett2018-engagement}.
Addressing these challenges requires more bespoke, careful design work (e.g.,~\cite{McCurdy2019-implicitError})
\revision{and a framework of ``open civic design''}~\cite{Reynante2021-openCivicDesign} \revision{to scaffold participation}.
The added labor of adopting such software \revision{practices}---\eg{} data entry, design, development, \revision{coordination,} maintenance---should not be shouldered by policy workers when researchers leave~\cite{correll2019ethical} nor should it be the permanent burden of graduate students attached to these projects~\cite{akbaba2023troubling}.

It would be a mistake for the visualization community to view PB merely as another application area for which to build dashboards. 
The work of building civic capacity weaves the very fabric of our society.
In the context of PB, this endeavor deserves and may even require the sustained involvement of visualization professionals, who can help to 
facilitate for municipal governments and their constituents the mutual support and accountability which 
\revision{enable}
``empowered participation'' 
~\cite{Fung2004-empowered}.
\looseness=-1

\subsection{Limitations}
\label{sec:limitations}
We do not claim to have provided an exhaustive perspective on the question of how visualization can benefit municipal governments, or even PB in Chicago. 
The most important omission from our investigation is the perspective of community members.
From the stance of feminist HCI~\cite{bardzell2011towards}, we note it is not ideal to study tools without including a major group of stakeholders who will use them.
Although we attempted to schedule focus groups with community members, we found it difficult to make connections representative of broader community interests in Chicago.
For example, we struggled to secure invitations to the community meetings used to mediate the PB process in Chicago. 
We speculate that organizers might not have wanted us to give community members the impression that something like our probe was on the horizon, especially given Chicago's unique history of civic inequity and fragility of the hard-won trust between ward offices and their constituents~\cite{Fung2004-empowered}.

By focusing on the perspective of PB experts, many of whom implement or study the PB process, our investigation provides an important view of how visualization and software more broadly fits into government procedures.
Prior work on PB tools has noted the need for work emphasizing this perspective~\cite{Harding2015-civicHCI-trust} because after all community members sometimes have poorly calibrated expectations about what is legal or feasible within the framework of government.

Future work should create visualization tools for PB through participatory design methods~\cite{spinuzzi2005-participatory}, and should investigate how the perspectives of PB organizers and residents come together in the context of deployment~\cite{Harding2015-civicHCI-trust}.
Doing so could corroborate, extend, and build greater ecological validity around some of our findings, e.g., about the fraught relationship between transparency and trust.
Additionally, we call for controlled experiments to address empirical questions raised here, \eg{} about how the provision of cost information and the design of value elicitation interfaces might influence stated preferences.

\subsection{\revision{Challenges for Visualization Research}}
\revision{Fulfilling the promise of visualization for PB and digital civics more broadly will require advancements across key topics of inquiry.}

\revision{\textbf{Persuasion and personalization.}
The use cases we envision for deliberation and storytelling require new ways of designing persuasive visualizations that are tailored to the context of a particular individual or group.
Residents and PB organizers would benefit from specialized visualization authoring tools for constructing their own data-driven arguments---e.g, using personalized annotations to highlight value conflicts in a community, or presenting evidence from historical data to argue for a particular course of action.
This aligns with recent calls to include decision-makers in designing visualizations of the criteria, alternatives, and preferences around decisions}~\cite{oral2023-visToolsForDM}.
\revision{Visualization specialists can provide support in crafting these designs but should avoid attempting to impose the illusion of a neutral ``view from nowhere''}~\cite{haraway2016situated}.
\revision{Along with persuasive tools, the visualization community must create ethical standards}~\cite{correll2019ethical} \revision{for how these tools can be used responsibly---e.g., outlining what kinds of design tactics are permissible in a voting interface.}
\looseness=-1

\revision{\textbf{Graphical literacy and accessibility.}
Visualization for digital civics requires developing more inclusive design practices. 
For example, the dashboard in our probe} (\figref{fig:map}, \figref{fig:strip-plot}) \revision{attempted to facilitate too many comparisons within one view, and replacing it with a more streamlined narrative presentation as we suggest would entail future work on the staging of multiple simpler views} (e.g.,~\cite{hullman2013-sequences}) \revision{and guidelines about what designs are appropriate at a given level of graphical literacy.
In developing visualizations to deploy for public audiences, future work should investigate ways of providing comparable affordances to people who are blind or live with other disabilities that impact the usefulness of visualization tools.
Translation across sensorimotor modalities may be especially challenging for graphical elicitation techniques, and for this reason we suggest that button-press and bubble-fill interactions may be preferable to direct manipulation for preference elicitation.}

\revision{\textbf{Uncertainty communication.}
Conveying uncertainty around political processes poses difficulties for PB tools and digital civics more broadly.
For example, our interviews show that uncertainty about how cost estimates are generated, or how votes are counted, triggers concerns about mistrust.
Discussions with participants raised possibilities around visualizing cost estimate distributions, or using sketchy rendering to signify that PB votes are not a final decision.
However, future work is needed to investigate the impact of these design choices on voting behavior and trust, as well as how to integrate uncertainty visualizations with explanations of political context.}

\revision{\textbf{Wading into politics.}}
As tool-builders, we must recognize how our design choices \revision{address political questions, such as one} raised by \pxx{01}, \pxx{12}, \revision{and} \pxx{13}: \emph{should PB be viewed more as a mechanism for public discussion or a decision-making procedure?}
We conclude that to treat PB only as a decision process begets excessive constraint and to treat it as mere discussion begets a lack of accountability; it must be both.
It would be impossible to design PB tools in a way that is agnostic to such political questions, and thus, 
\revision{visualization researchers} 
must play a humble role in defining the political processes we help to implement, although it may border on epistemic trespass in uncomfortable ways.

\pagebreak
\section{Supplementary Materials}
\label{sec:supp}
Our supplementary materials consists of a live demo of our design probe, the code for that probe, our interview guide, and the codebook from our qualitative analysis.
The probe can be found at \liveLink{}.
All other materials can be found at \osf{}.